\documentclass[12pt]{article}
\usepackage{amsmath}
\usepackage{epsfig}
\usepackage{verbatim}
\begin{document}
%
% For posting on archives
%
%\begin{comment} 
\begin{center}
\LARGE
\textbf{On relativistic elements of}\\
\textbf{reality}\\[1cm]
\large
\textbf{Louis Marchildon}\\[0.5cm]
\normalsize
D\'{e}partement de physique,
Universit\'{e} du Qu\'{e}bec,\\
Trois-Rivi\`{e}res, Qc.\ Canada G9A 5H7\\
email: marchild$\hspace{0.3em}a\hspace{-0.8em}
\bigcirc$uqtr.ca\\
\end{center}
\medskip
%
%\end{comment}
%
% For Foundations of Physics
%
\begin{comment}
\title{On relativistic elements of reality}

\author{Louis Marchildon}

\institute{L. Marchildon \at
              D\'{e}partement de physique,
              Universit\'{e} du Qu\'{e}bec,
              Trois-Rivi\`{e}res, Qc.\ Canada G9A~5H7\\
              Tel.: 819-376-5107\\
              Fax: 819-376-5164\\
              \email{louis.marchildon@uqtr.ca}}

\date{Received: date / Accepted: date}

\maketitle
\end{comment}
%
\begin{abstract}
Several arguments have been proposed some years
ago, attempting to prove the impossibility of
defining Lorentz-invariant elements of reality.
I find that a sufficient condition for the
existence of elements of reality, introduced
in these proofs, seems to be used also as a
necessary condition.  I argue that Lorentz-invariant
elements of reality can be defined but, as Vaidman
pointed out, they won't satisfy the so-called
product rule.  In so doing I obtain algebraic
constraints on elements of reality associated
with a maximal set of commuting Hermitian
operators.
%\keywords{Quantum mechanics \and Interpretation
%\and Lorentz invariance \and Elements of reality}
%\PACS{03.65.Ta \and 03.30.+p}
% \subclass{MSC code1 \and MSC code2 \and more}
\end{abstract}
%
%\newpage
\section{Introduction}
\label{sec1}
The notion of ``element of reality'' was introduced
in the famous Einstein, Podolsky and Rosen (EPR)
paper~\cite{einstein}, as an attribute of a physical
quantity whose value can be predicted with certainty
without disturbing the system.  Criticizing
the EPR conclusion that quantum mechanics is incomplete,
Bohr~\cite{bohr} argued that the phrase ``without in any
way disturbing a system'' is ambiguous and emphasized
the necessity of taking into account the experimental
arrangement with which a physical quantity is
measured.  To avoid the ambiguity, Redhead~\cite{redhead}
gave the following sufficient condition for the
existence of an element of reality, hereafter called ER1:
\begin{quote}
If we can predict with certainty, or at any rate with
probability one, the result of measuring a physical
quantity at time $t$, then at the time $t$ there
exists an element of reality corresponding
to the physical quantity and having a value equal to
the predicted measurement result. [\textbf{ER1}]
\end{quote}

A number of interpretations of quantum mechanics
involving various kinds of elements of reality were
proposed after the EPR paper, in particular
Bohmian mechanics~\cite{bohm} and modal
interpretations~\cite{vermaas}.  These were
originally developed as nonrelativistic theories,
and it has been notoriously difficult to reconcile
them with the special theory of relativity.
Eventually, the question was raised whether
Lorentz-invariant elements of reality are inconsistent
with quantum mechanics~\cite{hardy1,clifton1,clifton2}.

The purpose of this paper is to revisit that
question, and bring a number of additional
considerations to it.  I will first analyze in
detail Hardy's argument, which was meant to show
that Lorentz-invariant elements of reality are
indeed inconsistent with quantum
mechanics.\footnote{Hardy's paper also provides
a very interesting proof of Bell's theorem that
does not make use of inequalities.  That part of
the paper will not be discussed here.}  I will argue that
this and related arguments use ER1 not only as a
sufficient condition for the existence of an
element of reality, but also, it seems, as a
necessary condition.  A contradiction then
immediately follows, quite independently from
details of quantum mechanics.  My argument
will make use of constraints, derived in the
appendix, on elements of reality associated
with a maximal set of commuting Hermitian
operators.  I will then investigate to
what extent the light cone associated with an event
can be used to define Lorentz-invariant elements
of reality.  It turns out to be possible, but
these elements of reality won't satisfy
the so-called product rule, i.e. an element of
reality associated with a product of two commuting
operators will not always be equal to the product
of elements of reality associated with each
operator~\cite{vaidman1,vaidman2}.
Building on this I will discuss a number
of analyses of Hardy's and related
arguments published in the literature.
%
%\newpage
\section{Hardy's argument}
\label{sec2}
\begin{figure}[hbt]
\begin{center}
%\epsfbox{figure1.eps}
\epsfig{file=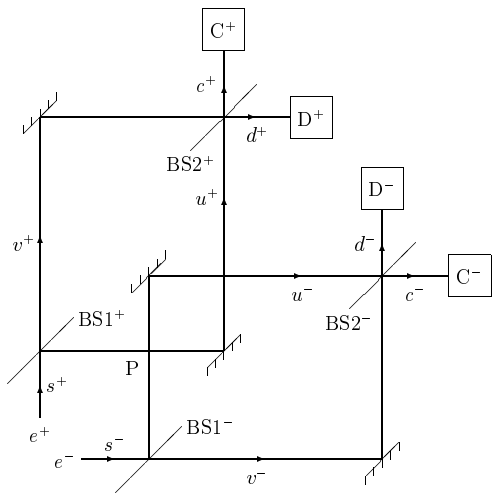,height=65mm,width=65mm}
\end{center}
\caption{Hardy's thought experiment with two
Mach-Zehnder-type interferometers}
\label{fig1}
\end{figure}

Hardy's gedanken experiment~\cite{hardy1}
is illustrated in Fig.~\ref{fig1}.
Two Mach-Zehnder-type interferometers are set up,
one for electrons (MZ$^-$, lower right) and one for
positrons (MZ$^+$, upper left).  Electrons (positrons)
are emitted with initial wave function $s^-$ ($s^+$).
Two beam splitters BS1$^{\pm}$ act in such a way that
\begin{equation}
|s^{\pm} \rangle \rightarrow \frac{1}{\sqrt{2}}
(i |u^{\pm} \rangle + |v^{\pm} \rangle) .
\label{bs1}
\end{equation}
The overlap of the two interferometers allows
paths $u^-$ and $u^+$ to meet at point $P$.  In that
case the electron and positron annihilate, i.e.
\begin{equation}
|u^{+} \rangle |u^{-} \rangle \rightarrow |\gamma \rangle .
\label{gamma}
\end{equation}

Let the initial state of the system be
$|s^{+} \rangle |s^{-} \rangle$.  Using (\ref{bs1})
and~(\ref{gamma}), we find that beyond beam
splitters BS1$^{\pm}$ and beyond point $P$, the
state of the system has been transformed into
\begin{equation}
\frac{1}{2} (\mbox{} - \gamma
+ i |u^{+} \rangle |v^{-} \rangle
+ i |v^{+} \rangle |u^{-} \rangle
+ |v^{+} \rangle |v^{-} \rangle ) .
\label{pointp}
\end{equation}

In each interferometer, the two paths are
reflected by mirrors arranged so that the paths
eventually meet again.  At the meeting points
are two additional beam splitters BS2$^-$ and
BS2$^+$ that act on the incoming beams so
that\footnote{In Ref.~\cite{hardy1}, beam
splitters BS2$^-$ and BS2$^+$ are removable,
but we don't need this flexibility here.}
\begin{equation}
|u^{\pm} \rangle \rightarrow \frac{1}{\sqrt{2}}
(|c^{\pm} \rangle + i |d^{\pm} \rangle)
\quad \mbox{and} \quad
|v^{\pm} \rangle \rightarrow \frac{1}{\sqrt{2}}
(i |c^{\pm} \rangle + |d^{\pm} \rangle) .
\label{BS2}
\end{equation}

Let $F$ be the rest frame of the two
interferometers.  We assume that in $F$, the 
electron and positron in each run of the experiment
reach BS2$^-$ and BS2$^+$ simultaneously.  Just before
this then, the state of the system is given
by~(\ref{pointp}).  We now
consider two additional Lorentz frames $F^-$ and
$F^+$, moving with respect to $F$ with opposite
velocities.  These velocities are chosen so that
in $F^-$, the electron arrives at BS2$^-$ much before
the positron arrives at BS2$^+$, whereas in $F^+$, the
positron arrives much before the electron.  Consider a time
in $F^-$ when the electron has gone through BS2$^-$,
but the positron has not yet reached BS2$^+$.  Using
(\ref{pointp}) and~(\ref{BS2}), we find that the
state of the system is then given by
\begin{equation}
\frac{1}{2 \sqrt{2}} (\mbox{} - \sqrt{2} \gamma
- |u^{+} \rangle |c^{-} \rangle
+ 2 i |v^{+} \rangle |c^{-} \rangle
+ i |u^{+} \rangle |d^{-} \rangle ) .
\label{Fminus}
\end{equation}
Similarly, at a time in $F^+$ when the positron has
gone through BS2$^+$ but the electron has not yet
reached BS2$^-$, the state of the system is given by
\begin{equation}
\frac{1}{2 \sqrt{2}} (\mbox{} - \sqrt{2} \gamma
- |c^{+} \rangle |u^{-} \rangle
+ 2 i |c^{+} \rangle |v^{-} \rangle
+ i |d^{+} \rangle |u^{-} \rangle ) .
\label{Fplus}
\end{equation}
Finally, we note that when both the electron and
positron have gone through the second beam splitters,
the state of the system is given by
\begin{equation}
\frac{1}{4} (\mbox{} - 2 \gamma
- 3 |c^{+} \rangle |c^{-} \rangle
+ i |c^{+} \rangle |d^{-} \rangle
+ i |d^{+} \rangle |c^{-} \rangle
- |d^{+} \rangle |d^{-} \rangle ) .
\label{after}
\end{equation}

To argue against relativistic elements of reality,
Hardy proposes a sufficient condition for their
existence and a necessary condition for their Lorentz
invariance.  The sufficient condition for the
existence of an element of reality essentially
coincides with Redhead's ER1.  The necessary
condition for Lorentz invariance of elements of
reality, hereafter called LI1, simply reads as:
\begin{quote}
The value of an element of reality corresponding to a
Lorentz-invariant observable is itself Lorentz
invariant. [\textbf{LI1}]
\end{quote}
I shall denote an element of reality
associated with an observable $A$ by $f(A)$.  Whenever
ER1 is satisfied for $A$, then
$f(A)$ coincides with an eigenvalue of $A$, a real number.

Hardy introduces two observables $U^{\pm}$ defined as
\begin{equation}
U^{\pm} = |u^{\pm} \rangle \langle u^{\pm} | .
\end{equation}
He then points out that
\begin{enumerate}
\item in state $|u^+\rangle$, $f(U^+) = 1$;
\item in state $|u^-\rangle$, $f(U^-) = 1$;
\item in state $|u^+\rangle |u^-\rangle$,
$f(U^+ U^-) = 1$;
\item in any state $|u^+, u^- \rangle_{\perp}$
orthogonal to $|u^+\rangle |u^-\rangle$,
$f(U^+ U^-) = 0$.
\end{enumerate}
From ER1 he concludes that 
\begin{equation}
f(U^+) f(U^-) = 1 \Rightarrow f(U^+ U^-) = 1 .
\label{FUU}
\end{equation}

We can now state Hardy's argument against
Lorentz-invariant elements of reality.  Consider
a situation where, in frame $F^+$, a positron is
detected in $D^+$ before the electron reaches BS2$^-$.
Then the system's state (\ref{Fplus}) is projected
onto its last term, and the electron's state becomes
$|u^-\rangle$.  In this case $f(U^-) = 1$ in $F^+$
and, by LI1 and
since $U^-$ is Lorentz invariant, $f(U^-) = 1$
in all Lorentz frames.  Similarly, consider
a situation where, in frame $F^-$, an electron is
detected in $D^-$ before the positron reaches BS2$^+$.
Then the system's state (\ref{Fminus}) is projected
onto its last term, the positron's state becomes
$|u^+\rangle$ and $f(U^+) = 1$ in all Lorentz frames. 
According to (\ref{after}), both these situations
will occur together, on average, in one of every
sixteen runs.  Whenever they occur
together, in all frames before
detection we have $f(U^-) = 1$ and $f(U^+) = 1$,
whence by (\ref{FUU}), $f(U^+ U^-) = 1$.  But
according to (\ref{pointp}) the state in $F$, the
rest frame of the two interferometers, is
orthogonal to $|u^+\rangle |u^-\rangle$.  Hence
$f(U^+ U^-) = 0$, which establishes the contradiction
and shows that Lorentz-invariant elements of
reality are incompatible with quantum mechanics.
%
%\newpage
\section{Analysis of the argument}
\label{sec3}
In formulating his argument, Hardy made explicit
use of the projection postulate.  One can see,
however, that this amounts to little more than a
verbal simplification, and that the use of the
postulate is in no way necessary to reach the intended
conclusion.  Indeed consider
the steps leading to the conclusion that
$f(U^-) = 1$ in $F^+$.  From (\ref{Fplus}), one easily sees
that the probability of obtaining the result~1 upon
measuring $U^-$, conditional on obtaining the
result~1 upon measuring $D^+$, is given by
\begin{equation}
P(U^- \rightarrow 1|D^+ \rightarrow 1)
= \frac{P(U^- \rightarrow 1 \mbox{ and } D^+ \rightarrow 1)}
{P(D^+ \rightarrow 1)}
= \frac{|i/(2\sqrt{2})|^2}{|i/(2\sqrt{2})|^2} = 1 .
\end{equation}
This means that if we obtain~1 upon measuring $D^+$,
we can predict with certainty that we will obtain~1
upon measuring $U^-$.  By ER1 then, and without any appeal
to the projection postulate, $f(U^-) = 1$.

Let us now focus on condition ER1.  Just like
Redhead, Hardy intends it as a sufficient condition
for the existence of an element of reality.  I
will now argue that it is actually used in a much
stronger way.  To see this, consider the following
steps of a possible proof of~(\ref{FUU}):
\begin{enumerate}
\item Assume that $f(U^+) f(U^-) = 1$.  Since $f(U^{\pm})$
is either 1 or~0, we must have both $f(U^+) = 1$
and $f(U^-) = 1$.
\item Since there is an element of reality corresponding
to $U^+$ having value~1, we can predict with certainty
that measuring $U^+$ will yield value~1.  Therefore
the system's state $|\Phi\rangle$ is an eigenstate
of $U^+$ with eigenvalue~1.  Similarly, $|\Phi\rangle$
is an eigenstate of $U^-$ with eigenvalue~1.
\item Since $|\Phi\rangle$ is an eigenstate of $U^+$
and $U^-$ with eigenvalue~1 in each case, it is also an
eigenstate of $U^+ U^-$ with eigenvalue~1.
\item Since $|\Phi\rangle$ is an eigenstate of $U^+ U^-$
with eigenvalue~1, we can predict with certainty that
measuring $U^+ U^-$ will yield value~1.  Hence
$f(U^+ U^-) = 1$.
\end{enumerate}

Looking closely at step~2, one can see that it assumes
that ER1 is not only a sufficient, but also a
necessary condition.  Admittedly, steps~1--4 represent
one possible reconstruction of the proof of~(\ref{FUU}),
and we cannot rule out other reconstructions that
would use weaker conditions.  But no proof of~(\ref{FUU})
is provided in Ref.~\cite{hardy1} assuming ER1 as a
sufficient condition only.

That ER1 is in fact used as a necessary
condition gains additional support from the
assumption that the elements of
reality obey the so-called product rule, i.e.
\begin{equation}
f(U^+ U^-) = f(U^+) f(U^-) .
\label{product}
\end{equation}
I show in the
appendix that any real-valued function $f$ which (i) is
defined on a maximal set of commuting Hermitian operators,
(ii) satisfies the product rule, and (iii) is 1 on some
but not all unit projectors in the set, singles out a
one-dimensional subspace of the state space, i.e.\ there
is only one projector in the set on which $f = 1$.
If the element of reality that this function assigns
is identified with an eigenvalue of a quantum
observable, then a unique state
vector is singled out by the specification that $f = 1$.
Hence it leads to the most definite predictions
that quantum mechanics allows.

One should note that the same analysis applies to the
proofs given in Ref.~\cite{clifton1} and~\cite{clifton2}.
In~\cite{clifton1}, an inference similar to~(\ref{FUU}) is
used at the bottom of the second column of p.~181.
In~\cite{clifton2}, the authors consider separately the
cases where QM is or is not complete.  In the former
(p.~124), elements of reality are taken to be eigenvalues,
which necessarily satisfy the product rule.  In the latter
(p.~125), the product rule is explicitly assumed.

On reading Hardy's and related arguments, one may
be tempted to question the appropriateness of conjoining
a manifestly non-Lorentz-invariant sufficient
condition for the existence of elements of reality
(ER1) with a requirement of Lorentz invariance of
some elements of reality (LI1).  Assuming that ER1
is used not only as a sufficient but also as a
necessary condition, the contradiction follows in
a strikingly simple way: (a) For the experiment
under consideration, the premise of ER1,
as this condition is formulated, is true in some frames
and not in others.  (b) LI1 says that
if the consequent of ER1 holds in one frame, then (for
Lorentz-invariant observables) it holds in all frames.
(c) But ER1 is also a necessary condition;
therefore its premise must hold in all frames.
We see that (c) plainly contradicts (a).
%
%\newpage
\section{Lorentz-invariant elements}
\label{sec4}
Condition ER1 involves the
word ``predict'' in an essential way.  In the technical
context in which it is used here, just like in its
normal usage, ``predict'' refers to the future.
The premise of ER1 assumes that we know
the (relevant) world configuration shortly before $t$
in one Lorentz frame, and can infer from this the result
of measuring the physical quantity $A$ at time~$t$.

Because ER1 refers to a specific Lorentz frame,
it is clearly not a relativistically
invariant criterion.  Indeed in the Hardy setup,
a prediction in frame $F^+$ can become a retrodiction
in $F^-$, and vice versa.  So how can we get an
invariant criterion?  The answer is that, instead
of specifying a hypersurface in a noninvariant way
(i.e. at constant time), we should use an invariant
specification.  The way to do this is to make
use of the light cone whose apex coincides with
the measurement event.\footnote{Of course a
measurement does not occur at a specific space-time
point, but I will assume that the argument can
be adapted to a more realistic situation.}
The light cone, however, can be used in a number
of different ways, which will have to be investigated
more closely.

Suppose we use the backward light cone of the
measurement event.  A sufficient criterion for
the existence of an element of reality could
then be the following:
\begin{quote}
If from the (relevant) information on or inside
the backward light cone of a possible measurement event
$\mathcal{E}$, we can infer with certainty,
or at any rate with probability one, the result
of measuring a physical quantity at
$\mathcal{E}$, then at that event there
exists an element of reality corresponding
to the physical quantity and having a value equal to
the predicted measurement result. [\textbf{ER2}]
\end{quote}

Formulated in this way,
ER2 is a perfectly valid and relativistically
invariant criterion.  It is,
however, completely inadequate as
a specification of elements of reality in the sense
of EPR.  Indeed ER2 can only serve to specify elements of
reality in the absolute future of the agent making the
inference.  But EPR had in mind elements of reality
whose values could, in some circumstances at least,
be inferred at events spacelike separated from the
agent.

So instead of using the backward light cone, we
should make use of the forward one.  Here then is
an appropriate sufficient criterion for the
existence of an element of reality:
\begin{quote}
If from the (relevant) information on or outside
the forward light cone of a possible measurement event
$\mathcal{E}$, we can infer with certainty,
or at any rate with probability one, the result
of measuring a physical quantity at
$\mathcal{E}$, then at that event there
exists an element of reality corresponding
to the physical quantity and having a value equal to
the predicted measurement result. [\textbf{ER3}]
\end{quote}

We should note that the word ``infer'' here can
cover both prediction and retrodiction.  In
ER2, it covers only prediction.
Readers who are familiar with the Hellwig-Kraus theory of
state vector collapse~\cite{hellwig} will probably
see the analogy between that theory and ER3.
In the Hellwig-Kraus approach, the collapse of the
state vector occurs on the past light cone of the
measurement event (say $\mathcal{M}$),
instead of on an equal time
hypersurface.  Such a collapse can therefore be
relevant to the attribution of an element of
reality to a physical quantity at an event
$\mathcal{E}$ that is spacelike separated
from (and in the relative past of)
the collapse triggering event $\mathcal{M}$.
I shall come back
to the Hellwig-Kraus approach in Sec.~\ref{sec5},
and argue that it stands in spite of objections
made to it in the literature.

\begin{figure}[hbt]
\begin{center}
%\epsfbox{figure2.eps}
\epsfig{file=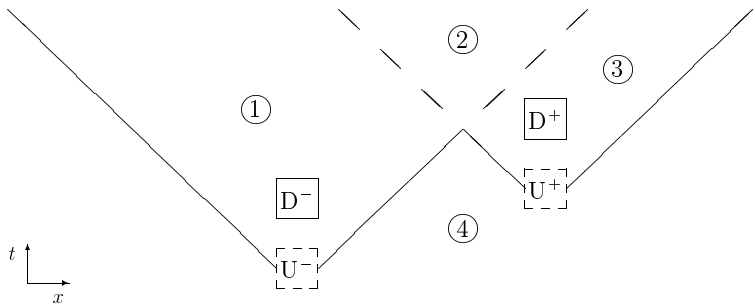,height=40mm,width=100mm}
\end{center}
\caption{Combination of two light cones}
\label{fig2}
\end{figure}

But we are not yet over.  ER3 is a suitable
condition for the existence of an element of reality
pertaining to a local physical quantity.  We
still have to adapt it to nonlocal quantities,
like $U^+ U^-$ in Hardy's setup.  There are several
obvious invariant
ways to combine two light cones, through their
union and their intersection.  These operations
are illustrated in Fig.~\ref{fig2}, for one space and one
time dimension.  We shall denote by $\mathcal{U}$
the union of the regions outside the forward light
cones of the boxes associated with $U^+$ and $U^-$.
Specifically, $\mathcal{U}$ contains regions~1, 3
and~4 in Fig.~\ref{fig2}.
Likewise we shall denote by $\mathcal{I}$ the
intersection of the regions outside the forward light
cones of the boxes.  $\mathcal{I}$
coincides with region~4 in Fig.~\ref{fig2}.

Suppose that we try to adapt criterion ER3 to
$\mathcal{U}$.  It is clear
that any time an actual measurement of $U^+ U^-$ is
performed, its result can be deduced from information
available in $\mathcal{U}$, for this
region includes part of the absolute future of the
measurement events.  Hence the criterion is in this
case trivial.  I shall come back in Sec.~\ref{sec5}
to the case where, as in Hardy's argument, no actual
measurement of $U^+ U^-$ is performed.

Suppose next that we try to adapt criterion ER3 to
$\mathcal{I}$.  Let us see what
we can say about elements of reality pertaining to
the physical quantities $U^+$, $U^-$ and $U^+ U^-$
in a run where detection occurs at $D^+$ and $D^-$.
Since $D^-$ is outside the forward light cone of
$U^+$, ER3 implies that $f(U^+) = 1$.  Likewise
since $D^+$ is outside the forward light cone of
$U^-$, ER3 implies that $f(U^-) = 1$.  This holds
in all Lorentz frames.

It turns out, however, that neither $D^+$ nor $D^-$
are in $\mathcal{I}$, the intersection of the
regions outside the two forward light cones.
Hence ER3 cannot be used to deduce the existence
of an element of reality associated with $U^+ U^-$,
nor \emph{a fortiori} to attribute a value to
$f(U^+ U^-)$.  Hardy's argument therefore no longer
goes through.

Thus if we pick $\mathcal{I}$ (i.e.\ region~4)
as the relevant region
for inferring elements of reality like $f(U^+ U^-)$,
we have $f(U^+) = 1 = f(U^-)$ but the value of
$f(U^+ U^-)$ cannot be inferred.  It therefore
follows that the product rule~(\ref{product}) no
longer necessarily holds.  This is the price we
have to pay for this notion of relativistically
invariant elements of reality.
%
%\newpage
\section{Discussion}
\label{sec5}
Hardy's and related arguments were criticized by
various authors, from several different angles.
I will here review these criticisms in the
light of the present analysis.

In a detailed analysis of Hardy's argument,
Clifton and Niemann~\cite{clifton1} suggested that
the Lorentz-invariance criterion LI1 is too strong.
They proposed instead the following:
\begin{quote}
If an element-of-reality corresponding to some
Lorentz invariant physical quantity exists and has
a value within a spacetime region R with respect
to one spacelike hyperplane containing R then,
\textit{if it exists with respect to another
hyperplane} H \textit{containing} R, it has
the same value in R with respect to H. [\textbf{LI2}]
\end{quote}
Using LI2 instead of LI1, one easily sees that
Hardy's argument no longer goes through.  If,
however, the objective is to have a Lorentz-invariant
theory at the level of elements of reality, then
LI2 will not meet it.  For LI2 suggests that
we will have elements of reality corresponding to
Lorentz-invariant physical quantities that will
exist in some Lorentz frames and not in others.

Another criticism of the argument, made in
Refs.~\cite{clifton1} and~\cite{berndl},
is that elements of reality are defined
without sufficiently taking context
into account.\footnote{Replying to~\cite{berndl},
Hardy~\cite{hardy2} conceded that if one does not
assume locality, his argument doesn't hold.
In so doing he seems to have viewed condition ER1
differently.}  It is well known that in
theories involving hidden variables, their
values depend on context, for instance on what
specific observable is being measured far away.
In Hardy's setup, one can measure different
observables related to
the positron and the electron, by
removing the mirrors BS2$^+$ and BS2$^-$.
However, Cohen and Hiley~\cite{cohen1} pointed out
that to carry out the argument, Hardy only uses
a single experimental context, the one
where both mirrors are in place
and both detectors $D^+$ and $D^-$ fire.
Thus context dependence construed as above
is not relevant to the analysis.  On the other
hand, the criterion ER3 for the existence of
elements of reality, and its generalization to
nonlocal observables through the region
$\mathcal{I}$ introduced in Sec.~\ref{sec4},
do in a quite specific sense involve context.
Indeed the ``relevant information'' is not the
same for the local observable $U^+$ as it is for
the nonlocal observable $U^+ U^-$.  The expression
``context dependence'' may therefore remain more
appropriate than the expression ``observer specific''
suggested in~\cite{cohen1}.

Cohen and Hiley have also argued that the use of the
projection postulate is the crux of the problem.
In Sec.~\ref{sec3}, however, we showed that although
Hardy formulated his argument in the language
of the projection postulate, it doesn't really
depend on it.  By using a functional-based
formalism for the description of the state of
a quantum-mechanical system, Cohen and Hiley can
provide a relativistically-covariant description
of whatever collapses occur in the Hardy
experiment.  That formalism, however, doesn't
seem to allow for elements of reality to be
introduced even when measurement results can
be predicted with certainty.\footnote{Dewdney and
Horton~\cite{dewdney} have introduced
Lorentz-invariant elements of reality associated
with Bohmian trajectories, but their approach
appears to involve significant constraints on
the particles' motion.  A Lorentz-invariant
realistic approach to quantum mechanics was also
proposed by Tumulka~\cite{tumulka} in the context of
the Ghirardi-Rimini-Weber model of spontaneous
collapse, again with restrictions on the particles'
interaction.  Ref.~\cite{tumulka} also provides a brief
overview of the literature on relativistic models
explaining quantum-mechanical probabilities.}

Vaidman~\cite{vaidman1,vaidman2} has argued that
insistence on the validity of the product
rule~(\ref{product}), on which Hardy's and other
arguments are based, is what prevents defining
Lorentz-invariant elements of reality.  To show
this, Vaidman applied the Aharonov-Bergmann-Lebowitz
(ABL) rule to the setup.  In its original
formulation~\cite{ABL}, the rule gives the
probability $\mbox{Prob}
(P_i \rightarrow 1 |\phi, \psi)$ that,
if a system is prepared in a state $|\phi\rangle$
and found in a final stage to be in a state
$|\psi\rangle$, an intermediate measurement
associated with a projector $P_i$ will find the
result~1:
\begin{equation}
\mbox{Prob}(P_i \rightarrow 1 |\phi, \psi)
= \frac{\mbox{Tr} (P_{\psi} P_i P_{\phi} P_i)}
{\sum_j \mbox{Tr} (P_{\psi} P_j P_{\phi} P_j)} .
\end{equation}
Here $P_{\phi}$ and $P_{\psi}$ project on states
$|\phi\rangle$ and $|\psi\rangle$.
The sum in the denominator runs on a set of
orthogonal projectors summing to the
identity.\footnote{If the system's Hamiltonian
doesn't vanish between measurements, $|\phi\rangle$
must be evolved forwards to the time of the
intermediate measurement and $|\psi\rangle$
must similarly be evolved backwards.}

Vaidman then shows that if $|\phi\rangle$ is the
state given in~(\ref{pointp}) and $|\psi\rangle
= |d^+\rangle d^- \rangle$ (i.e. the state
obtained upon detection by $D^+$ and $D^-$),
then there is unit probability for the
following three intermediate results: (i)
$U^+$ yields 1; (ii) $U^-$ yields 1; and (iii)
$U^+ U^-$ yields 0.

These probabilities are unobjectionable if
the intermediate measurements are carried out.
But in Hardy's experiment, they are not.
However, Vaidman extrapolates the ABL rule
to counterfactual situations.  He proposes to
attribute elements of reality to physical quantities
whose values can be deduced with certainty,
on the basis of a counterfactual use of the
ABL rule.  From the results given above,
it is clear that these elements of reality
do not satisfy the product rule.

Can functions defined on commuting Hermitian
operators and not satisfying the product rule
still be called elements of \emph{reality}?
This is a semantical question that I shall not
address, but there is no doubt that, if anything,
they make for a peculiar kind of reality.  Cohen
and Hiley~\cite{cohen2} have argued that the
violation of the product rule entails a meaning
of \emph{true} which ``differs from its meaning
in standard predictive quantum mechanics and in
all other branches of physics.''  But the difference
may not be so drastic.  A proposition $p$ and its
negation may both be true if they refer to different
times, i.e.\ to different contexts.  As we have
seen above, elements of reality are in some sense
context dependent.  Note that an operational,
albeit indirect, meaning to the violation of the
product rule can in the present situation be provided
by weak measurements~\cite{aharonov2}.

Cohen and Hiley also objected to Vaidman's
counterfactual use of the ABL rule, on the grounds
that the observables involved do not form the
basis of a consistent family of histories.
It would seem, however, that this lack of a
consistent family only prevents assigning to
elements of reality values that can be revealed
in and would be unaffected by an eventual
measurement.  Much of the debate on the counterfactual
use of the ABL rule comes from the failure to
realize that a number of different assertions can
be made about counterfactual measurement results
that are all consistent with the formalism of
quantum mechanics.  I have argued
elsewhere~\cite{marchildon} that the
counterfactual use of the ABL rule, in situations
where it is allowed, is not a consequence
of the formalism of quantum mechanics, but
is part of its interpretation.

I should point out that Vaidman's analysis
corresponds to the choice indicated in
Sec.~\ref{sec4}, where the value of the element
of reality corresponding to the nonlocal
observable $U^+ U^-$ could be
deduced from information in the union
$\mathcal{U}$ of the regions outside the two
forward light cones of the events.  As he claims,
it thus provides Lorentz-invariant elements
of reality.

In closing this section, I would like to argue
for the consistency of the Hellwig-Kraus
relativistic theory of the projection postulate,
against the criticism made to it by Aharonov
and Albert~\cite{aharonov}.

Briefly, Aharonov and Albert correctly
pointed out that some nonlocal observables
can be measured by means of local measurements
only.  To illustrate this, they considered
two spin 1/2 particles prepared in the
singlet state and then separated at an
arbitrarily large distance.  Next they introduced
two measurement apparatus, correlated in such
a way that the difference of their
(three-dimensional) positions and the sum of
their momenta are well defined.  They then
showed that these apparatus can be used, say
at $t = - \epsilon$ in some frame, to assert
that each component of the two particles' total
spin vanishes, in a non-demolition way.

Now suppose that at time $t$ in that frame, the
$z$-component of the spin of the left particle is
measured.  According to the Hellwig-Kraus theory,
that particle's state then collapses onto
$|+; \hat{z} \rangle$ or
$|-; \hat{z} \rangle$ along the backward
light cone of the particle.  This, Aharonov and
Albert maintain, contradicts the fact that the
two-particle state is $|S=0\rangle$ just before
the measurement (at $t = - \epsilon$).

According to the Hellwig-Kraus theory, however,
the local non-demolition measurements at
$t = - \epsilon$ will not collapse the nonlocal
state on this equal-time hypersurface (that would
not be relativistically invariant), but on the
union of the two particles' backward light cones.
Hence the two-particle state
$|S=0\rangle$ will be valid in the
region bounded by these light cones and the one
coming from the measurement event of the $z$-spin
of the left particle.
%
%\newpage
\section{Conclusion}
\label{sec6}
Hardy's and other proofs of the nonexistence of
Lorentz-invariant elements of reality rely in one
way or another on the assumption that the element
of reality associated with a product of commuting
Hermitian operators is the product of the elements
of reality associated with each individual operator.
This assumption, as shown in the appendix, constrains
values of elements of reality defined on a maximal
set of commuting Hermitian operators.
Once it is removed, Lorentz-invariant
elements of reality can be defined using various
combinations of the light cones associated with
measurement events.  This construction, although
not necessarily involving state vector collapse,
is reminiscent of the Hellwig-Kraus theory.
The kind of context dependence involved allows
for the failure of the product rule and clarifies
the sense in which these elements of reality
differ from what this concept usually covers.
%
%\begin{acknowledgements}
\section*{Acknowledgements}
I am grateful to the Natural Science and Engineering
Research Council of Canada for financial support.
It is a pleasure to thank Simon
L\'{e}vesque for discussions on a related topic
which stimulated some of the thoughts
presented here. My thanks also go to Andrew Laidlaw
for comments, and to two anonymous referees for
insightful remarks.
%\end{acknowledgements}
%
%\newpage
\section*{Appendix}
In this section I address the question of
finding real-valued functions, on a maximal set
of commuting Hermitian operators, that satisfy
the product rule.

Let $\mathcal{V}$ be a complex vector space of
finite dimension $N > 2$, and let $\{|u_i\rangle \}$
be an orthonormal basis in $\mathcal{V}$.  Then the
set of all operators of the form
\begin{equation}
H = \sum_{i=1}^N \lambda_i |u_i\rangle \langle u_i | ,
\label{hermitian}
\end{equation}
with the $\lambda_i$ real, is a maximal set of commuting
Hermitian operators.  We look for real functions
$f(H)$ such that the product rule is satisfied, i.e.
\begin{equation}
f(H_a) f(H_b) = f(H_a H_b)
\label{product2}
\end{equation}
for any $H_a$ and $H_b$ having the form (\ref{hermitian}).

For any projector $P$, (\ref{product2}) implies that
$f(P) = f(P P) = f(P) f(P)$, so that $f(P) = 0$ or~1.
Let $P_i$ denote the one-dimensional projector
$|u_i\rangle \langle u_i |$.  All possible cases
then fall into the following three disjoint collections:
\begin{enumerate}
\item $f(P_i) = 1$ for all $P_i$.
\item $f(P_i) = 1$ for some $P_i$ and 0 for others.
\item $f(P_i) = 0$ for all $P_i$.
\end{enumerate}
Let us now examine these possibilities in more
detail.
\paragraph{Case 1}
Let $K = P_i + \lambda_j P_j$ with $i \neq j$ and
$\lambda_j \neq 0$.  Then
\begin{equation}
f(K) = f(K) f(P_i) = f(K P_i) = f(P_i) = 1
\end{equation}
and
\begin{equation}
f(K) = f(K) f(P_j) = f(K P_j) = f(\lambda_j P_j) ,
\end{equation}
whence $f(\lambda_j P_j) = 1$ for all $j$ and $\lambda_j
\neq 0$.  For any nontrivial $H$ as given 
in~(\ref{hermitian}), there is a $k$ such that
$\lambda_k \neq 0$.  But then  
\begin{equation}
f(H) = f(H) f(P_k) = f(H P_k) = f(\lambda_k P_k) = 1 .
\end{equation}
Note that since $P_i P_j = 0$ we also have $f(0) = 1$,
hence $f(H) = 1$ for all $H$.
\paragraph{Case 2}
Here there has to be exactly one $P_i$ such that
$f(P_i) = 1$.  For if $f(P_i) = 1 = f(P_j)$ with
$i \neq j$, then
\begin{equation}
f(0) = f(P_i P_j) = f(P_i) f(P_j) = 1 .
\label{fzero}
\end{equation}
But if $k$ is such that $f(P_k) = 0$,
\begin{equation}
f(0) = f(P_i P_k) = f(P_i) f(P_k) = 0 ,
\label{f0}
\end{equation}
which contradicts (\ref{fzero}).

Now let $H$ be as in~(\ref{hermitian}).  Then 
\begin{equation}
f(H) = f(H) f(P_i) = f(H P_i) = f(\lambda_i P_i) .
\label{fh}
\end{equation}
Moreover
\begin{equation}
f(\lambda_i P_i) f(\lambda'_i P_i)
= f(\lambda_i \lambda'_i P_i) .
\label{flambda}
\end{equation}
Restricted to positive values of $\lambda_i$,
(\ref{flambda}) means that $f$ is a one-dimensional
representation of the Lie group of products of
real numbers.  In the exponential parametrization
$\lambda = \exp (\ln \lambda)$, hence representations
have the form $f(\lambda_i P_i) = \exp (\alpha \ln \lambda_i)
= (\lambda_i)^{\alpha}$, where $\alpha$ must be real
for $f$ to be.  To extend $f$ to negative values of
$\lambda_i$ we can either put
$f(- |\lambda_i| P_i) = |\lambda_i|^{\alpha}$ for
all $\lambda_i$ or
$f(- |\lambda_i| P_i) = - |\lambda_i|^{\alpha}$ for
all $\lambda_i$.  Finally, for $f$ to remain finite as
$\lambda_i \rightarrow 0$, the exponent
$\alpha$ must be nonnegative.  This also follows
from~(\ref{f0}) if $f$ is required to be continuous.
It then follows from~(\ref{fh}) that for all $H$
\begin{equation}
f(H) = |\lambda_i|^{\alpha}
\end{equation}
or
\begin{equation}
f(H) = \begin{cases}
(\lambda_i)^{\alpha} & \text{if } \lambda_i \ge 0 , \\
- |\lambda_i|^{\alpha} & \text{if } \lambda_i < 0 . \\
\end{cases}
\label{twodim}
\end{equation}
\paragraph{Case 3}
We see at once that $f(0) = 0$.  It may be that there
is a two-dimensional projector $P_{ij} = P_i + P_j$
such that $f(P_{ij}) = 1$.  Then for any projector
which is a sum of one or several $P_k$
orthogonal to $P_{ij}$,
\begin{equation}
f(P) = f(P_{ij}) f(P) = f(P_{ij} P) = f(0) = 0 
\end{equation}
and
\begin{equation}
f(P + P_i) = f(P + P_i) f(P_{ij})
= f[(P + P_i) P_{ij}] = f(P_i) = 0 .
\end{equation}
Likewise we easily see that $f(P + P_j) = 0$
and $f(P + P_{ij}) = 1$.  If we let
$f(H) = |\lambda_i |^{\alpha} |\lambda_j |^{\beta}$
for $\alpha > 0$ and $\beta > 0$, we find at once that
(\ref{product2}) is satisfied for all $H_a$ and $H_b$.
It can also be satisfied with various choices similar
to (\ref{twodim}).

If there is no two-dimensional projector $P_{ij}$
such that $f(P_{ij}) = 1$ (and therefore
$f(P_{ij}) = 0$ for all $i$ and $j$), it may be that there is
a three-dimensional projector $P_{ijk} = P_i + P_j
+ P_k$ such that $f(P_{ijk}) = 1$.  Letting
$f(H) = |\lambda_i |^{\alpha} |\lambda_j |^{\beta}
|\lambda_k |^{\gamma}$ (with $\alpha > 0$,
$\beta > 0$ and $\gamma > 0$), we see that (\ref{product2})
is satisfied for all $H_a$ and $H_b$.  The same argument
goes on in higher dimensions.  It is tempting to
conjecture that this exhausts all possibilities,
but I do not have a proof of this general
statement.
%
%\newpage

%
\end{document}